# Vertical interface enabled large tunability of scattering spectra in lightweight microwire/silicone rubber composites

A. Uddin, D. Estevez, H.X. Peng and F.X. Qin[*]

*Institute for Composites Science Innovation (InCSI), School of Materials Science and Engineering, Zhejiang University, 38 Zheda Road, Hangzhou, 310027, PR. China*

**Abstract**

Previously, we have shown the advantages of an approach based on microstructural modulation of the functional phase and topology of periodically arranged elements to program wave scattering in ferromagnetic microwire composites. However, the possibility of making full use of composite intrinsic structure was not exploited. In this work, we implement the concept of material plainification by an in-built vertical interface on randomly dispersed short-cut microwire composites allowing the adjustment of electromagnetic properties to a large extent. Such interface was modified through arranging wires of different structures in two separated regions and by enlarging or reducing these regions through wire concentration variations leading to polarization differences across the interface and hence microwave tunability. When the wire concentration was equal in both regions, two well-defined transmission windows with varied amplitude and bandwidth were generated. Wire concentration fluctuations resulted in strong scattering changes ranging from broad passbands to stopbands with pronounced transmission dips, demonstrating the intimate relationship between wire content and space charge variations at the interface. Overall, this study provides a novel method to rationally exploit interfacial effects in microwire composites. Moreover, the advantages of enabling significantly tunable scattering spectra by merely 0.053 vol. % filler loading and relatively simple structure make the proposed composite plainification strategy instrumental to designing microwave filters with broadband frequency selectivity.

---

[*] corresponding author: faxiangqin@zju.edu.cn

# 1. Introduction

Composites containing ferromagnetic wires have attracted much attention owing to their multiple outstanding properties and associated broad range of applications such as structural health monitoring, microwave absorption and electromagnetic shielding [1–5]. Strong responses from the microwires to the incident electromagnetic wave are possible due to their distinguished giant magneto- and stress-impedance effects, soft magnetic character and emerging properties arising from their arrangement in the matrix [4,6–8]. It is possible to further tune the scattering spectra of microwire composites through external stimuli such as magnetic field, temperature or stress but these approaches entail practical burdens, limiting their applications in compact and light-weight devices.

Beyond those methods, we have recently proposed a programming-based strategy using the wire (microstructurally modified by current annealing) as a code and different wire combinations as the code pattern [9,10]. Although such approach enabled modulation of scattering properties to a large extent, strategies other than increasing the number of components and modifying microstructure have not yet been exploited. We have previously demonstrated the novelty of designing vertical interfaces on carbon nanocomposites to trigger significant alteration of dielectric properties and hence scattering properties [11]. Space charges accumulate near the interface and by restraining their local distribution distinct dielectric responses can be obtained. This kind of so-called "plainified" composites design makes full use of the composite structure for maximizing specific performance while maintaining same filler content. Alloyed materials have also taken advantage of such approach by altering grain boundaries with fewer or zero alloying elements [12].

In this work, we move forward to apply these two strategies simultaneously to composites incorporating randomly dispersed short-wires. By arranging the as-cast and annealed wires in two regions of the composite, an additional factor influencing the wave propagation comes into play, i.e., interface-induced polarization. This polarization originated from the differences in conductivity, permittivity and relaxation time of charge carriers across the in-built vertical interface [11]. We then designed several representative vertical-interface composite systems by enlarging or reducing one of the regions through variations in the corresponding wire volume concentration. Scattering spectra having band-stop and band-pass regions with frequency and bandwidth tuning flexibility could be achieved through filler loadings as low as 0.053 vol. %. Therefore, a design method based on single-wire modification and the concept of plainification by in-built interfaces is sufficient to provide a broad scope of electromagnetic signatures. This could have substantial implications for the design of efficient and low-cost band-stop/band-pass filters. Moreover, the frequency shift could be conveniently used in practical applications such as stress monitoring, whereby the transformation of the scattering spectra will be similar but due to the tensile stress. These findings offer new opportunities of materials plainification design for developing large scale composites having novel properties without compromising lightweight feature or structural efficiency and integrity.

## 2. Experimental Method

*2.1 Fabrication and current annealing of $Co_{60}Fe_{15}Si_{10}B_{15}$ glass-coated microwires*

$Co_{60}Fe_{15}Si_{10}B_{15}$ glass-coated microwires with a total wire diameter ($D_w$) of 35.4 µm and metallic core diameter ($d_m$) of 27.2 µm were fabricated by a modified Taylor-Ulitovskiy technique [13]. Such fabrication method allows the production of continuous wires with easily controllable dimensions [13,14]. The produced microwires were current-annealed under different DC currents of 30 mA and 40 mA for 10 minutes by mechanically removing the glass coat from the wire ends to allow electrical contact. The changes in DC resistance of the current-annealed microwires were monitored with a digital multi-meter showing a considerable drop for the 40 mA treated-wire which is related to the onset of crystallization and larger grain size due to the rise in temperature [9,15]. The microstructure of the as-cast and current-annealed wires consists mainly of nanocrystalline droplets embedded in amorphous matrix [16]. Magnetic properties of the microwires were evaluated by Quantum Design PPMS-VSM at room temperature.

*2.2. Preparation and electromagnetic characterization of randomly dispersed short-cut $Co_{60}Fe_{15}Si_{10}B_{15}$ glass-coated microwires/silicone rubber composites*

Short-cut microwires of about 5 mm length were randomly dispersed in the matrix. SYLGARD(R) 184 silicone elastomer kit (Dow Corning) was used as the matrix containing base and curing agent which were thoroughly mixed using a weight ratio of 10:1. The silicone was poured into a designed cuboid mold and cured partially before the placement of the wires to avoid them from sinking to the bottom. Composite samples with dimensions of 22.86 x 10.16 x 2 mm$^3$ were fabricated after curing at 125 °C for 20 minutes. Two kinds of composites were prepared. The first type incorporating 50 wires (0.053 vol. %) of the same type, i.e. as-cast A, 30 mA-current annealed B or 40 mA-current annealed C (Fig. 1a). The second type corresponds to incorporating the as-cast A and annealed short-cut wires X (where X corresponds to B or C annealed wires) in separated regions and variable proportion in the matrix (Fig. 1b and Table. 1).

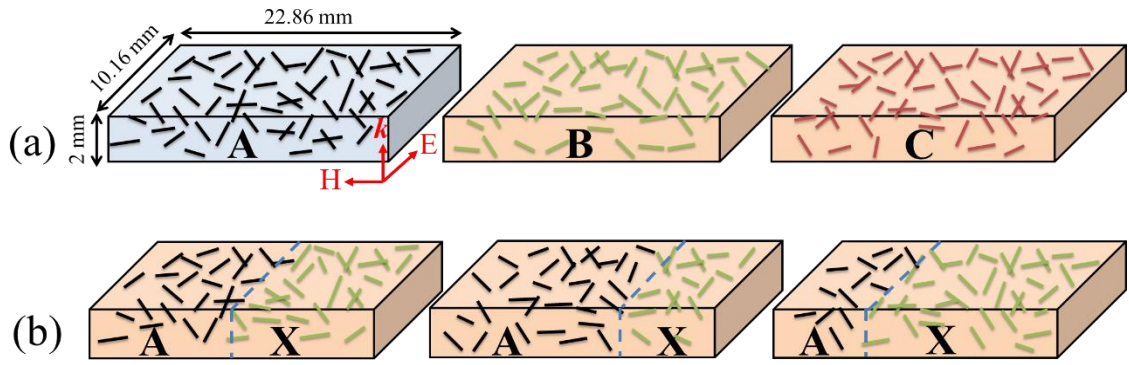

**Figure 1:** Randomly dispersed short-cut microwire composites. (a) Composites incorporating the same type of wire, as-cast A, 30 mA-annealed B or 40 mA-annealed C with a wire amount of 0.053 vol. %. (b) Composites combining the as-cast A and annealed wires X (where X corresponds to B or C annealed wires) in separated regions and different proportions. The dotted line represents the vertical interface formed between the two different regions.

**Table. 1:** Filler proportion in the composites combining as-cast A and annealed short-cut wires X (B: 30 mA or C: 40 mA) distributed in separated regions.

| As-cast Wire-A amount (vol. %) | Annealed Wires-X amount (vol. %) | | Total amount (vol. %) |
|---|---|---|---|
| | Wire-B | Wire-C | |
| 25 (0.026) | 25 (0.026) | - | 50 (0.053) |
| 25 (0.026) | - | 25 (0.026) | 50 (0.053) |
| 33 (0.035) | 17 (0.018) | - | 50 (0.053) |
| 33 (0.035) | - | 17 (0.018) | 50 (0.053) |
| 17 (0.018) | 33 (0.035) | - | 50 (0.053) |
| 17 (0.018) | - | 33 (0.035) | 50 (0.053) |

Scattering S-parameters of the composites were measured with Rohde & Schwarz ZNB 40 vector network analyzer (VNA) by using a WR-90 waveguide in $TE_{10}$ dominant mode from 8.2 to 12.4 GHz (Fig. 2). Before the measurements, the VNA was calibrated by the TRL (thru-reflect-line) calibration method [17].

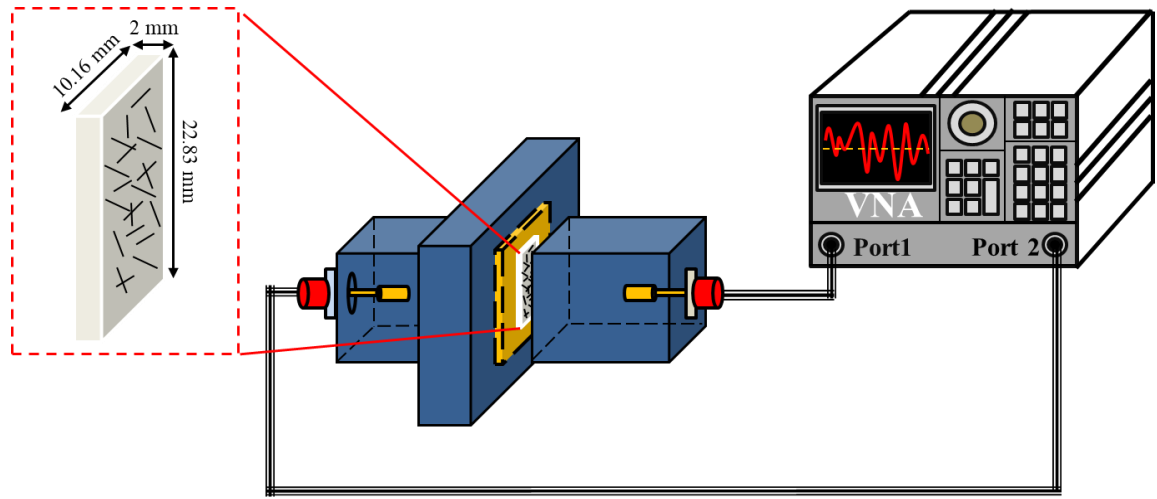

**Figure 2:** Measurement setup for obtaining the scattering parameters of the microwire composite sample. The waveguide system includes two coaxial adapters (Port 1 and 2) connected to a vector network analyzer (frequency range: 8.2–12.4 GHz). The size of the sample measured in the waveguide was $22.86 \times 10.16 \times 2$ mm$^3$.

### 3. Results & Discussions

*3.1. Magnetic and electrical properties of as-cast and current annealed $Co_{60}Fe_{15}Si_{10}B_{15}$ glass-coated microwires*

Figure 3 shows the magnetic hysteresis loops, magneto-impedance and electrical conductivity for the as-cast and current annealed microwires. Current annealing improves the magnetic softness evidenced by the drastic decrease in coercivity from 15 Oe for the as-cast wire to 0.8 Oe for the 40 mA-annealed wire and by the drop in magnetic anisotropy (Fig. 3a). From the two branches of the magneto-impedance curves (under increasing and decreasing magnetic fields) at 200 MHz (Fig. 3b), the as-cast wire is characterized by two peaks of weak intensity, which evidences a system with a weak circumferential magnetic anisotropy [18,19]. Current annealing the wire at 30 mA results in pronounced asymmetrical double peaks and lower impedance, demonstrating the gradual contribution to the magnetization of the circular domains at the wire surface shell [18,20,21]. Further annealing at 40 mA seems to deteriorate the magnetization along the circumference likely due to the above-mentioned onset of crystallization in that sample. Such changes in magnetization and impedance curves are related to relief of internal stresses and structural relaxation [22,23]. Changes in electrical conductivity also reflect the structural modification of the microwires during current annealing (Fig. 3c). The increase in conductivity in the 40 mA-annealed wire originates from the larger grain size, which is caused by Joule heating leading to less electron scattering at the reduced grain boundaries [9,24]. The structural changes due to annealing as indicated in both magnetic and electrical properties of the microwires will have a considerable influence on the electromagnetic properties of their composites, as will be discussed in the following section.

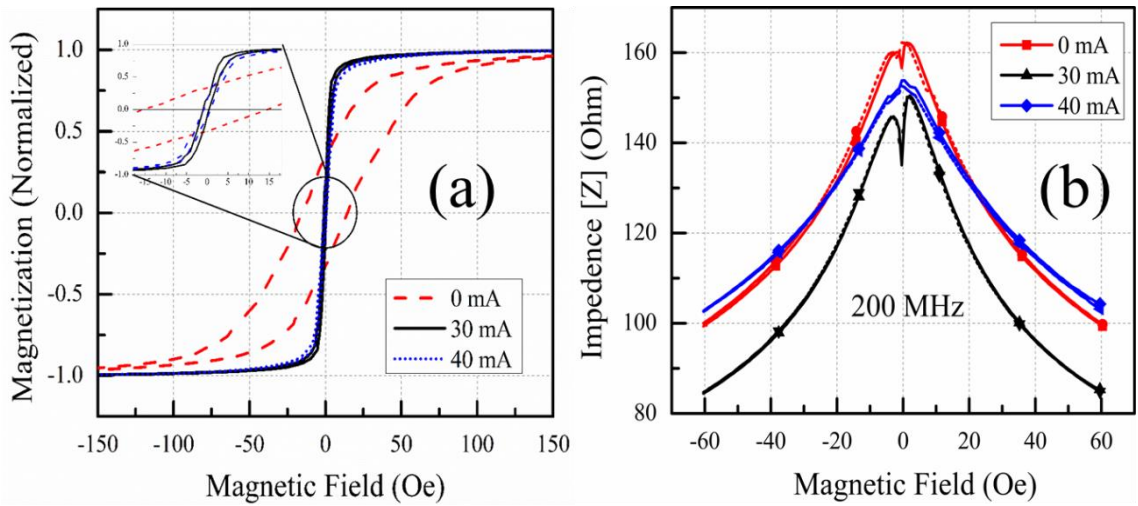

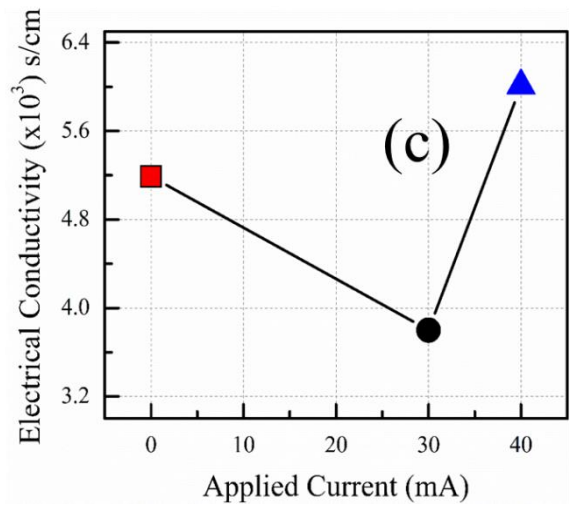

**Figure 3:** (a) Magnetic hysteresis M-H loops, (b) Magneto-Impedance at 200 MHz and (c) Electrical properties of as-cast, 30 mA and 40 mA current annealed $Co_{60}Fe_{15}Si_{10}B_{15}$ microwires. Inset in Fig. 3a shows the zoomed-in M-H loops at low field.

*3.2. Scattering properties of randomly dispersed short-cut $Co_{60}Fe_{15}Si_{10}B_{15}$ glass-coated microwires /silicone rubber composites*

The parameters that describe the material properties in microwaves frequencies are found from the measurement of the transmission $S_{21}$ and reflection $S_{11}$ coefficients. Figure 4 presents the scattering properties of the composites containing randomly dispersed microwires of the same type and a combination of as-cast and annealed wires. For the same type-wire composites, a passband extending through the whole range of measured frequencies appears in the transmission spectrum of composites containing as-cast wires A (Fig. 4a). This band-pass filter allows such specific band of frequencies to pass through while rejects the rest. When the inclusions are replaced by the annealed microwires, a switching from passband to stopband feature with transmission dip or resonance and associated reflection and absorption peaks is identified. Band-stop filters block frequencies that lie between its two cut-off frequency points ($f_L$ and $f_H$) but passes all those frequencies either side of this range. At the resonance, the current distribution in wires strongly depends on their surface impedance and, in general, the lower the surface impedance, the greater the scattering properties [20]. In microwires, the surface impedance reaches a minimum when the magnetization is along the circular direction, which corresponds to a circumferential anisotropy. This is confirmed by hysteresis loops of the annealed microwires (Fig. 3a) presenting almost linear shape with low coercivity and anisotropy field as a result of Joule heating. Thus, a deep minimum in transmission is seen for the composites containing the annealed short-cut microwires.

Moreover, short pieces of microwires interact with the electromagnetic radiation acting as micro-antennas. Thus, the wire length *l* and permittivity of the dielectric matrix $\varepsilon_m$ define the characteristic frequency related to the antenna resonance:

$$f_{res} = c/2l\sqrt{\varepsilon_m} \qquad (1),$$

where *c* corresponds to the speed of light [7]. For *l* =5 mm and $\varepsilon_m$= 2, the resonance frequency is close to 20 GHz, which falls out of the measurement range of the present work. The discrepancy between the theoretical value of the resonance and the one measured for the random-wire composites can be attributed to the fact that the long-range dipolar-dipolar interaction gives rise to additional anisotropies [25,26]. Therefore, at higher volume concentration, the antenna resonance equation is no longer applicable [27]. When the wires are overlapped with one another, the wires form clusters and the resonance frequency will be brought down to lower frequency [28]. The overlapping could be also responsible for the additional resonances seen for the composites containing the 30 mA annealed wires B [27]. Finally, the loading effect of the Pyrex glass coating should also be taken into account. The typical permittivity of Pyrex glass (4~10) is larger than that of silicone (~2), the resonance frequency is consequently shifted down [28].

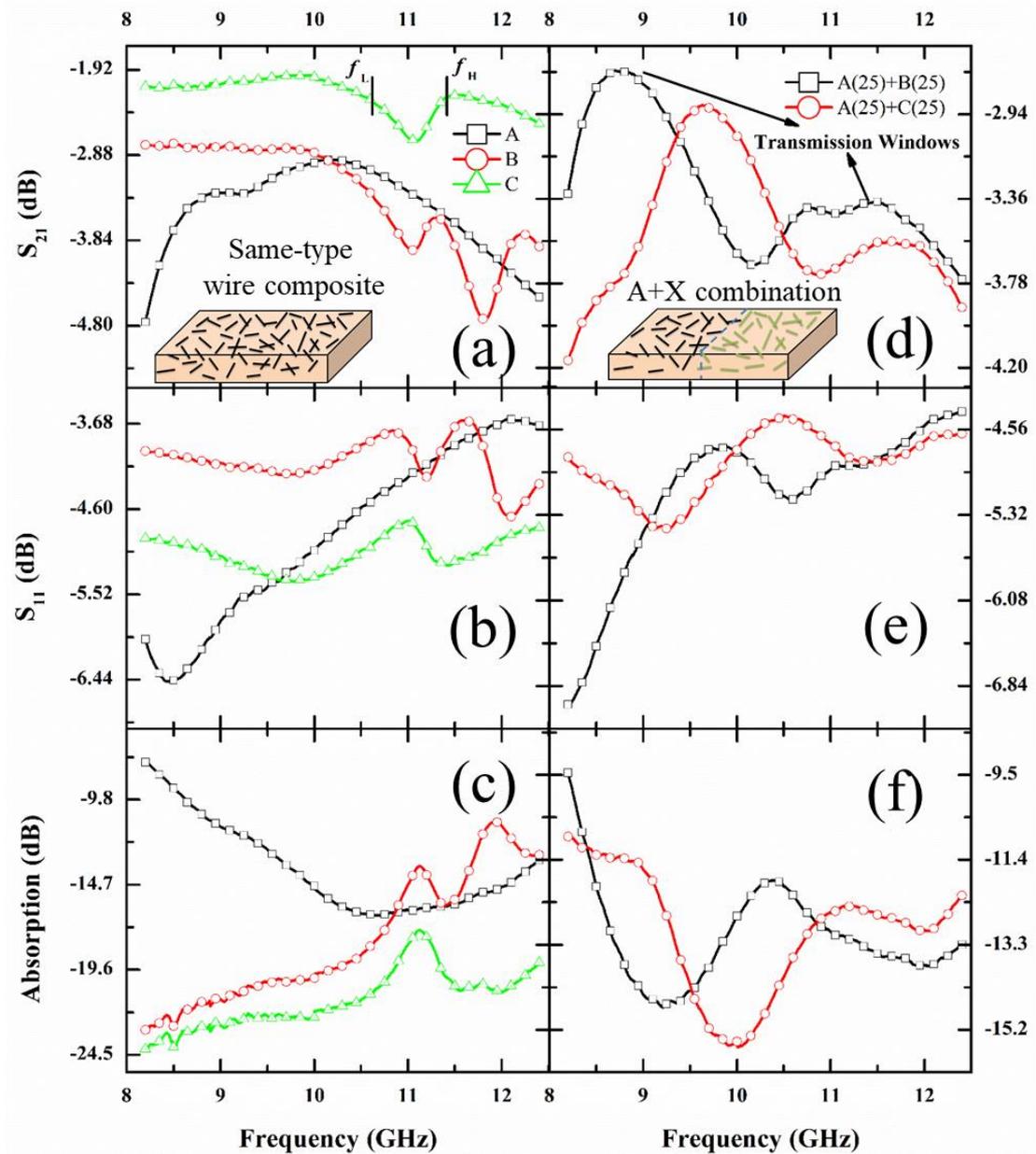

**Figure 4:** Scattering properties: transmission $S_{21}$, reflection $S_{11}$, and absorption of the composites containing same type (left panel) and an equal combination of as-cast A and annealed X (right panel) randomly dispersed short-cut microwires. The labels $f_L$ and $f_H$ correspond to the two cut-off frequency points of the stopband feature.

In order to exploit the concept of plainification in programming wave scattering of microwire composites, we built several vertical-interface composites with varied compositions. With respect to the scattering properties of composites containing an combination of equal loading of as-cast A and annealed wires X (Fig. 4, right panel), two transmission windows are observed together with dips in reflection and absorption spectra. The wires behave as a transmitting structure with band-pass feature in that condition. The transmission windows might be originated from the coexistence of the separated regions in the composite. Due to the differences in conductivity and permittivity between the left and right regions of the sample, the vertical interface becomes the center of charge accumulation under the electric field. These space charges modify the field distribution and lead to an additional term contributing to the overall dielectric properties of the composite and hence the electromagnetic response. Moreover, the transmission spectra move almost as a single unit to higher frequencies with lower magnitude as the 30 mA annealed wires (B) are replaced by the 40 mA annealed wires (C). Such changes are related to the gradual relief of internal stresses and induced structural relaxation experienced during the annealing process.

We move forward to tune the scattering properties of the composites by moving one region of the sample relative to another through variation in wire volume fraction (Fig. 1b). Incorporating majority of as-cast wires with respect to annealed wires (33 wires to 17 wires, respectively), results in distorted broad passbands and associated reflection and absorption peaks across the whole frequency range. In this case, the electromagnetic response is influenced mainly by the as-cast wires that contribute to the appearance of the passbands and by the interface between the two regions, which results in the splitting of the transmission spectra. When more annealed wires than as-cast wires (17 wires to 33 wires, respectively) are present, the composites behave as a stopband structure showing the dominant contribution of annealed wires to the electromagnetic signature. Similar to the case of composites containing only annealed wires B (Fig. 4a), dual stopbands are noticed in the A(17)+B(33) composites but with broader bandwidth. The single stopband or transmission dip found in composites containing only annealed wires C (Fig. 4a) is also displayed in the A(17)+C(33) composites, but a much deeper transmission minimum is present. It follows that the tunable band features can be modulated by varying the vertical interfacial region.

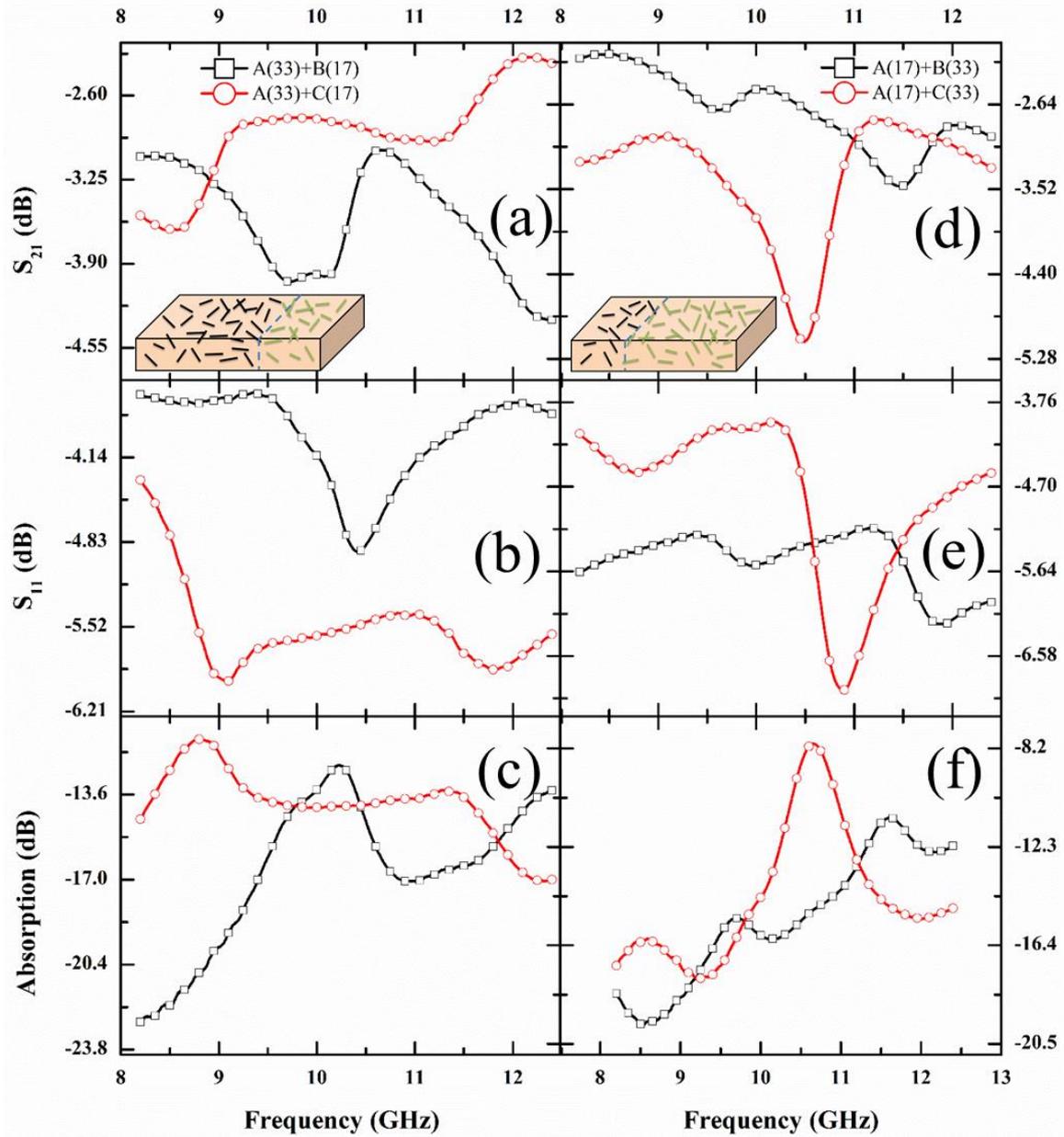

**Figure 5:** Scattering properties: transmission $S_{21}$, reflection $S_{11}$ and absorption of the composites containing different combinations of as-cast A and annealed X (30 mA, B or 40 mA, C) randomly

dispersed short-cut microwires in separated regions and different proportions, i.e., 33 wires of as-cast and 17 wires of annealed wires (right panel) and 17 as-cast and 33 annealed wires (left panel).

Now let us discuss the mechanisms behind the scattering response of the studied short-cut randomly dispersed composites. Note that all the composites under discussion are non-percolating due to the existence of glass coating and thus, there is no concern that the formation of a conductive network will hinder their interactions with the microwave [29]. The electromagnetic wave is scattered by inhomogeneities in the medium, and when such inhomogeneities are randomly distributed, the wave undergoes a large number of scattering events at random positions (Fig. 6a). Different regimes of scattering are noted depending on the degree of randomness, ranging from multiple scattering to strong wave localization in which the waves tend to be "localized" or "trapped" in the random media leading to reduced transmission [30,31]. The attenuation of the incident wave by the microwires occurs through loss phenomena originated from dielectric and magnetic losses, which can be controlled by changing the surface impedance and magnetic anisotropy in the wires [20,32]. From Kittel's equation [33], the ferromagnetic resonance frequency $f_{rm}$ is expressed as:

$$f_{r.m} = \frac{g}{2\pi}\sqrt{(H_a + 4\pi M)(H_a + H)} \qquad (2),$$

where $g$ is the gyromagnetic ratio, $H_a$ is the magneto-elastic anisotropy field which depends on the wire's dimensions, $M$ is the magnetization and $H$ is the applied external magnetic field. Therefore, apart from the length of the wires, the resonance frequency is also controlled by changes in internal anisotropy of the wires as per Eqs. (1) and (2). In the present case, such quantities were modified through current annealing as demonstrated by the hysteresis loops and impedance response (*Section 3.1*).

There are in general a number of critical parameters that control the microwave response of the studied composites, such as intrinsic properties of the wires, wire concentration and the in-built interface between the two different regions. For composites incorporating the same type of wires (Fig. 6a), the main factor influencing the scattering corresponds to the structure of the wires after the annealing process. As current annealing lowered the wire surface impedance (Fig. 3b), the scattering at the resonance increased and the rate of transmission was low [20]. However, as mentioned before, due to the closeness of crystallization and slightly larger anisotropy for the 40 mA-annealed wires, the surface impedance increased leading to higher transmission (Fig. 4a). In the case of composites including the separated regions, apart from the structure of the wires, the role of the introduced vertical interface and wire concentration on wave propagation should also be considered. Electromagnetic waves are generally strongly localized near the interface of the two different media (Fig. 6b). When the incident wave hits the interface between the two different media with unequal dielectric and magnetic properties, the amount of wave transmitted or reflected varies with the polarization state at the vertical interface. Each composite system is labelled from 1 to 6, representing different states of polarization between the two regions across the interface ($\Delta\varepsilon$) triggered by wire composition and filler microstructure variations (Fig. S1). Various dielectric dispersion behaviors are shown in the permittivity plots, ranging from almost constant response (systems 1 and 3) to relaxation (systems 2, 4 and 5) and resonance type (system 6). It is revealed that, in addition to wire conductivity and impedance [20,34,35], the introduced vertical interface plays a major role in formulating the electromagnetic patterns. It follows that, through controlling the composition in each region, the discrepancy between their permittivity (the ability to restrict and accumulate charges) is varied, giving rise to disparate interactions between the composites and electromagnetic waves. In general, fluctuations in density increase the apparent attenuation due to scattering because of increased impedance contrasts [36,37].

Our approach demonstrates new opportunities for developing advanced microwire composites by combining both strategies: microstructure modulation and plainification. Tunable properties in traditional composites are achieved here by simple structure, constant filler content and tailored interfaces at a different length scale. For instance, unlike magnetic field-tuned scattering behavior achieved in previous studies [38–40], tunable microwave band-stop and band-pass behaviors are realized without external magnetic fields. This eases the application of the studied composites in compact devices or light-weight structures where extra magnets are considered a burden. It also shows the potential of extrapolating the material plainification concept often used in nanostructures to large scale systems and realize superior performance without compromising lightweight design and structural integrity. Importantly, the passband and stopband performance with desired bandwidths could be further improved by hybridization with other functional fillers, inclusions alignment and supporting dielectric substrate modification. Relevant work is underway.

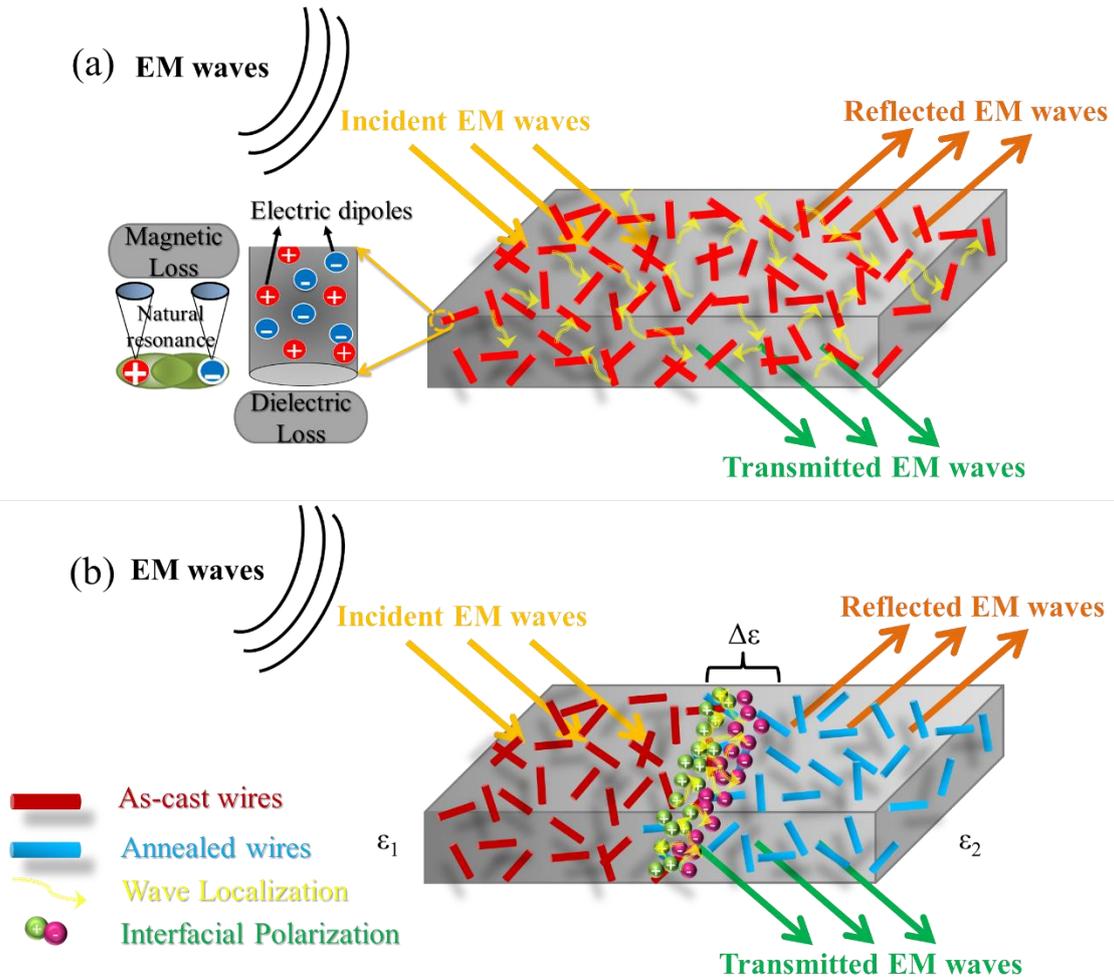

**Figure 6:** Proposed scattering mechanisms in the randomly dispersed short-cut microwire composites: (a) Composites containing the same type of wire; (b) Composites containing as-cast wires and annealed wires in separated regions.

## 4. Conclusions

We have successfully realized tunable microwave passband and stopband features in randomly dispersed wire composites through wire microstructure and vertical interfacial region modulation based on a plainification design philosophy. Structural relaxation of the current annealed wires reflected in the change of magnetic, impedance and electrical properties which in turn influenced the scattering properties of the wire composites. For the composites incorporating the same type of wire, a marked transition from passband to stopband behavior was shown when replacing as-cast wires by annealed wires due to refinement of circumferential anisotropy and lower impedance after annealing treatment. When a vertical interface was generated from incorporating different structural wires in separated regions, double transmission windows with different amplitudes were obtained. The differences in conductivity and permittivity of charge carriers across the interface resulted in an interface-induced polarization which affected the wave propagation characteristics in the composites. Such vertical interface was modified by enlarging or reducing one of the regions through variations in wire volume concentration, enabling further tunable scattering spectra from broad passbands to stopbands with variable intensity. The advantages of large microwave tunability with low and constant filler loading and a relatively simple structure make the proposed composites attractive for the design of band-stop and band-pass filters with wideband frequency selectivity.

**CRediT authorship contribution statement**

**A. Uddin:** Conceptualization, Methodology, Validation, Formal analysis, Investigation, Data curation, Writing - original draft. **D. Estevez:** Conceptualization, Validation, Formal analysis, Investigation, Writing - review & editing. **H. X. Peng:** Writing - review & editing. **F.X. Qin:** Conceptualization, Methodology, Validation, Formal analysis, Resources, Supervision, Project administration, Funding acquisition.

**Acknowledgments**

This work was supported by NSFC No. 51701178, No. 51671171 and ZJNSF No. LR20E010001 and No. LY20E010005, Peng is indebted to the support by the Fundamental Research Funds for the Central Universities; Qin is also indebted to the support of the "National Youth Thousand Talents Program" of China.